\documentclass[prl,twocolumn,showpacs]{revtex4}
\usepackage{graphics,amsmath,amssymb}

\newcommand{\pao}{\ensuremath{\psi_{a1}} }
\newcommand{\pbo}{\ensuremath{\psi_{b1}} }
\newcommand{\pat}{\ensuremath{\psi_{a2}} }
\newcommand{\pbt}{\ensuremath{\psi_{b2}} }
\newcommand{\pa}{\ensuremath{\hat\psi_{a}(z,t)} }
\newcommand{\pad}{\ensuremath{\hat\psi_{a}^\dagger(z,t)} }
\newcommand{\pb}{\ensuremath{\hat\psi_{b}(z,t)} }
\newcommand{\pbd}{\ensuremath{\hat\psi_{b}^\dagger(z,t)} }

\begin{document} 

\title{Squeezed light from spin squeezed atoms}

\author{Uffe V. Poulsen}
\email{uvp@ifa.au.dk}
\author{Klaus M{\o}lmer}
\affiliation{ Institute of Physics and Astronomy, University of Aarhus,
    DK-8000 \AA rhus C, Denmark}


\begin{abstract}
  We propose to produce pulses of strongly squeezed light
  by Raman scattering of a strong laser pulse on a spin squeezed
  atomic sample. We prove that the emission is restricted to a single
  field mode which perfectly inherits the quantum correlations
  of the atomic system.
\end{abstract}  

\maketitle

Squeezed light and entangled beams of light can be used to probe
matter and to study mechanical motion with better resolution than
classical light \cite{squeezing_review}.  Entangled photon sources can
be used for lithography with a resolution below the optical wavelength
\cite{holography,bjork01:_sub}, and the active field of quantum
information profits from the development of non-classical light
sources \cite{mattle96, bouwmeester97, furusawa98}.  Non-linear
crystals in optical parameter oscillators and laser diodes with
suitable feed-back have been the work horses in most experiments on
non-classical light.  As a figure of merit for the degree of
non-classicality of these light sources, one may refer to the noise
suppression observed in direct or homodyne photon detection
measurements. Compared to classical sources it has so far been
possible to reduce the noise (variance) by about one order of
magnitude. In order to make a significant difference in practical
applications, further noise reduction is really necessary.

It has been proposed that the large non-linearity and low absorption
in resonant Raman systems can lead to ideal four-wave mixing and
substantial squeezing~\cite{hemmer95, lukin98, lukin99}. As an
alternative approach we propose in this Letter to use spin squeezed
ensembles of atoms as sources of squeezed light. Atoms can be
entangled in such a way that the fluctuations in occupancies of
different internal states are significantly suppressed.  This
phenomenon is refered to as spin squeezing, because a two level atom
can be described formally as a spin 1/2 particle, and the interest in
spin squeezed states arose already a long time ago in connection with
ultra-precise spectroscopy and atomic
clocks~\cite{wineland94:_squeez}.  Squeezing of spins was originally
believed to be very complicated, but recent proposals based on quantum
non-demolition measurements of atomic populations \cite{kuzmich}, on
coherent interactions in Bose-Einstein
condensates~\cite{soerensen01:_many_bose, poulsen01:_posit_p_bose},
and on interactions between laser excited atoms \cite{bouchoule01}
have changed this impression and suggested that really significant
spin squeezing is achievable.  The main purpose of the Letter is to
demonstrate that the atomic quantum correlations can be perfectly
transferred to the field. This result is readily obtained within a
simplified model where both atoms and field are described by single
harmonic oscillators~\cite{jing01} but we show that correlations in
the atoms can be mapped perfectly on the field also in an a priori
multimode situation.

\begin{figure}[p] 
  \begin{center} \resizebox{8.5cm}{!}{
        \includegraphics{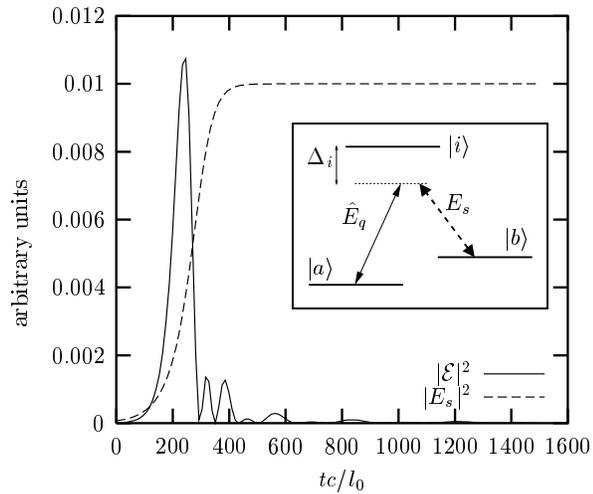}  }
\caption{The shape of the $E_s$
pulse and the mode function $\mathcal{E}$ of the emitted $\hat{E}_q$
pulse. The mode function is found by diagonalizing the first-order
correlation function of the field, but coincides with the signal
expected when the single atom operator $\hat{Q}$ is uniform over the
sample. The insert shows the atomic level scheme of the proposal.} 
  \label{fig:signal}
\end{center}
\end{figure}

The emission of light is treated by a simple generalization of the
theory of stimulated Raman scattering \cite{raymer81:_stimul_raman,
raymer85:_quant_raman,carman70:_theor_stokes} to and this part of our
proposal can be analyzed without specifying the model for spin
squeezing.  Our ensemble of two-state atoms is assumed to be strongly
elongated and it is treated in a 1D approximation. It is illuminated
by a strong laser field $E_s$ propagating along the z-axis of the
system.  This opens up a channel for an atom in the $b$ state to go to
the $a$ state by absorbing a photon of frequency $\omega_s$ and
emitting a photon of frequency $\omega_q\sim \omega_s+\omega_{ba}$. As
a result, a field at this frequency builds up and propagates through
the sample.  This process is described by the following coupled set of
equations for the atomic and field operators
\begin{gather}
  \begin{split}
  \label{eq:eom_pbdpa}
    \frac{\partial}{\partial t} & \left[ \pbd  \pa \right] 
        = 
     -i  \kappa_1^*  E_s^*(z,t) \times \\
    & \times \left( \pad \pa - \pbd \pb \right) 
  \hat{E}_q(z,t)
  \end{split}    
         \\
   \label{eq:eom_eq_pbdpa}
   \left(\frac{\partial}{\partial z} +
\frac{1}{c}\frac{\partial}{\partial t}\right)
   \hat{E}_q(z,t) =
   -i \kappa_2 \pbd \pa E_s(z,t).
\end{gather}
where $\kappa_1 = \sum_i \mu_{ai} \mu_{bi}/(\hbar^2\Delta_i)$ and
$\kappa_2 = 2\pi\hbar\omega_q\kappa_1/c$. $\mu_{ji}$ are dipole moments of the
atomic transitions and $\Delta_i$ are the (large) detunings with respect to
intermediate levels, see Fig.~\ref{fig:signal}.
$\pa,\ \pad$ and $\pb,\ \pbd$ are annihilation and
creation operators for atoms in states $a$ and $b$; $\pbd\pa$ is the 
positive frequency part of the atomic dipole operator, taking into account
the atomic density at position $z$ in the ensemble.
In Eq.~(\ref{eq:eom_pbdpa}) we have assumed  that there is no
dephasing of the $ab$ coherence and we have assumed the validity of
the slowly varying envelope approximation for the emitted field
in Eq.~(\ref{eq:eom_eq_pbdpa}).

We restrict our analysis to the case where the atoms are almost entirely in
the $a$ state when the $E_s$ field is applied. This implies  that the
population difference appearing in Eq.~(\ref{eq:eom_pbdpa}) 
can be replaced by the density of atoms $n(z)$. This density
we represent as a c-number throughout the duration of the output coupling.
This allows us to define a dipole operator "per atom" by
$\pbd\pa=n(z)\hat{Q}(z,t)$ and we obtain the linear operator equations
\begin{eqnarray}
  \label{eq:eom_q}
  \frac{\partial}{\partial t} \hat{Q}(z,t)
  &=&
  -i \kappa_1^* E_s^*(z,t) \hat{E}_q(z,t) 
  \\
  \label{eq:eom_eq}
  \left(\frac{\partial}{\partial z} + \frac{1}{c}\frac{\partial}{t}\right)
  \hat{E}_q(z,t) &=&
  -i \kappa_2 n(z) E_s(z,t) \hat{Q}(z,t)
  .
\end{eqnarray}

If we generalize the analysis in~\cite{raymer81:_stimul_raman,
raymer85:_quant_raman, carman70:_theor_stokes} to
inhomogeneous media, we can solve
Eqs.~(\ref{eq:eom_q}) and (\ref{eq:eom_eq}) analytically
in terms of the input field $\hat{E}_q(0,t)$ at the entrance face of
the sample at $z=0$ and the initial position dependent atomic 
polarization $\hat{Q}(z,0)$. 
In particular, for the field operator we get
\begin{multline}
  \label{eq:eq_of_tau}
  \hat{E}_q(z,\tau) = \hat{E}_q(0,\tau) 
  - i \kappa_2 E_s(\tau) 
  \int_0^z \! 
  \psi_b^\dagger(z',0) \psi_a(z',0) \times \\
  \times J_0\left(2\sqrt{a(\tau)\int_{z'}^z n(z'') dz'' }\right)
  \; dz'
\end{multline}
where the new time coordinate is $\tau\equiv t-z/c$, $J_0(\cdot)$ is a
Bessel function of the first kind, and
$a(\tau)=\kappa_1^*\kappa_2\int_0^\tau \left|E_s(\tau')\right|^2 d\tau'$. 

The expression (\ref{eq:eq_of_tau}) must be evaluated at the position
$z=L$ of a detector outside the atomic sample. In the absence of atoms,
the field equals the incident quantum vacuum field $\hat{E}_q(0,\tau)$.   
The atomic sample is able to replace the vacuum with an entirely different
field.  In order to analyze the quantum properties of $\hat{E}_q$ it is
convenient to imagine a time integrated homodyne dection at the
detector. By choosing
the temporal form of the strong local oscillator field in this detection 
we select a certain spatio-temporal mode of the field
represented by the field operator
\begin{equation}
  \label{eq:def_a}
  \hat{a}=\sqrt{\frac{c}{2\pi\hbar\omega_q}}
  \int_0^\infty
  \mathcal{E}^*(\tau)\hat{E}_q(L,\tau) 
  d\tau
\end{equation}
where $\int |\mathcal{E}|^2 d\tau = 1$. In choosing
$\mathcal{E}(\tau)$ we should seek to ensure that $\hat{a}$ is a
mapping of the precise collective operator of the atomic sample that
is known to be squeezed.  Mathematically, it is easy to show that in
order to probe $\int \! h(z') \hat{Q}(z',0) dz'$ we should choose
$\mathcal{E}(\tau)$ as the normalized solution of
Eqs.~(\ref{eq:eom_q},\ref{eq:eom_eq}) with the initial condition
$n(z')\hat{Q}(z',0)$ replaced by $h(z')$.  Physically, this
reflects that the question of mode matching coincides with the problem
of identifying the classical field radiated by a classical dipole
distribution. In particular, if the uniform integral of atomic
operators $\int \psi_b^\dagger(z',0) \psi_a(z',0) dz'$ is squeezed,
mode matching is accomplished by taking $\mathcal{E}(\tau)$ to be the
(normalized) solution to the classical Raman scattering problem.  In
that case we get the mapping
\begin{equation}
\label{eq:adagger_jplus}
\hat{a} = \frac{1}{\sqrt{N}}J_-
\end{equation}
where $\hat{J}_-\equiv \int \pbd\pa dz$,  and $N$ is the total number
of atoms.

It is a natural concern, whether a 3D analysis will preserve the
possibility to couple to only a single spatio-temporal field mode.
The issue has been addressed in the classical case, where a
diagonalization of the first order coherence function (Karhunen-Loeve
transformation) of the field indeed shows that a single mode dominates
the output, provided the Fresnel number of
the (active part of the) atomic sample is of the order of
unity~\cite{raymer85:_quant_raman}. Choosing the experimental
parameters accordingly we thus expect this result to apply also for
the quantum field in three dimensions.

We recall that several schemes now exist for the generation of
non-classical states of atomic spins.  The above analytical treatment
is general and independent of the method of spin squeezing, and as
shown by Eq.(7) (valid only if the majority of atoms occupy the state
$a$), the atomic state is transfered perfectly to the light
field. Mean values and variances for the field observables are
therefore explicitly known, and, {\it e.g.}, the orders of magnitude
squeezing derived in \cite{soerensen01:_many_bose,
poulsen01:_posit_p_bose} apply to the emitted light pulse. It may be
useful, however, to relate the emitted quantum field directly to the
dynamical variables in the spin squeezing process. This may provide
further insight in the origin of squeezing; it may provide a useful
tool for the application of the squeezed light as input to another
quantum system \cite{gardiner_cascade_prl, gardiner94}; and, it may be
useful to analyze processes where the spin squeezing and the light
emission occurs simultaneously.

As an example, consider spin 
squeezing by collisional interactions in a two-component  Bose Einstein
condensate \cite{soerensen01:_many_bose, poulsen01:_posit_p_bose}:
First, a Bose Einstein condensate is formed in only one of the internal
states $a$. By a short resonant Raman pulse, all atoms are transfered to an
equal superposition of $a$ and $b$.  
We assume that the interaction strengths between the atoms are not all 
equal, $g_{aa}=g_{bb}\ne g_{ab}$, so that the interaction terms in the
fully quantized interaction Hamiltonian
\begin{multline}
\label{full-ham}
    \hat{H} =
        \\ 
    \int \! d^3 r \Bigl\{
     \sum_{i=a,b} \Bigl[
      \hat{\psi}^{\dagger}_i(\vec{r}) \hat h_i
      \hat{\psi}(\vec{r})
   + \frac{g_{ii}}{2}
      \hat{\psi}^{\dagger}_i(\vec{r})
      \hat{\psi}^{\dagger}_i(\vec{r}) 
      \hat{\psi}_i(\vec{r}) \hat{\psi}_i(\vec{r}) 
    \Bigl]
    \\
    + g_{ab}       
    \hat{\psi}^{\dagger}_a(\vec{r})
    \hat{\psi}^{\dagger}_b(\vec{r}) 
    \hat{\psi}_b(\vec{r}) \hat{\psi}_a(\vec{r}) 
    \Bigl\}
\end{multline}
cause a Kerr-like phase evolution of amplitudes on states with different
numbers of atoms in the two internal states. This results in 
squeezing of an appropriate collective spin variable as pointed out by
S{\o}rensen et al.\ \cite{soerensen01:_many_bose}.  

As in our earlier work on this problem~\cite{poulsen01:_posit_p_bose}
we use the {\it positive $P$} method~\cite{gardiner91:_quant_noise,
steel98:_dynam_bose_einst, carusotto01:_n} to describe the squeezing
process.  This means that averages of all normal ordered operator
products can be calculated as ensemble averages  by 
the substitution of atomic field
operators $\hat{\boldsymbol{\psi}}\equiv
(\hat{\psi}_a,\hat{\psi}_a^\dagger,
\hat{\psi}_b,\hat{\psi}_b^\dagger)$ by pairs of
two-component ``wavefunctions'',
$\boldsymbol{\psi}=(\pao,\pat,\pbo,\pbt)$.  The dynamics of the wave
functions is given by four coupled and noisy ``Gross-Pitaevski''
equations, see details in Ref.~\cite{poulsen01:_posit_p_bose}.
Starting from a coherent initial state of all atoms in an equal
superposition of internal states $a$ and $b$, we numerically simulate
solutions of these equations to obtain an ensemble of
$\boldsymbol{\psi}$'s describing exactly (up to sampling errors) the
quantum correlations of the system.

Once a sizable spin squeezing is obtained we want to transfer this
special quantum state to a light pulse, and to be able to use the
results of the above analysis we 
first bring the internal state of the atoms close to the $a$ state, i.e.,
with a new resonant Raman pulse we rotate the collective spin close 
to the north pole of the Bloch sphere. This rotation is applied to the
individual sets of "wavefunction" realisations of the simulation
($i=1,2$):
\begin{equation}
  \label{eq:rot_pihalf}
   \left(
     \begin{array}{c}
       \psi_{ai} \\
       \psi_{bi}
     \end{array}
   \right)
   \rightarrow
   \left(
     \begin{array}{c}
       \psi_{ai}' \\
       \psi_{bi}'
     \end{array}
   \right)
   = 
   \left(
     \begin{array}{cc}
       \!\!\!\phantom{+}\cos\frac{\theta}{2} & \sin\frac{\theta}{2} \\
       \!\!\!-\sin\frac{\theta}{2} & \cos\frac{\theta}{2} 
     \end{array}
   \right)
   \left(
     \begin{array}{c}
       \psi_{ai} \\
       \psi_{bi}
     \end{array}
   \right)
\end{equation}
with $\theta\cong\pi/2$. 

The atom field interaction in the coupled equations~(\ref{eq:eom_pbdpa})
and (\ref{eq:eom_eq_pbdpa}) leads to a natural positive 
P representation of the
field $\hat{E}_q(z,t)$: Replace in Eq.~(\ref{eq:eq_of_tau})
$\hat{E}_q(z,t)$ by a c-number field $E_{q1}(z,t)$ and
$\psi_b^\dagger(z',0) \psi_a(z',0)$ by $\psi_{b2}(z',0)
\psi_{a1}(z',0)$, and make a similar replacement in the hermitian
conjugate equation where $\psi_a^\dagger(z',0) \psi_b(z',0)$ is
replaced by $\psi_{a2}(z',0) \psi_{b1}(z',0)$ to yield the c-number
field $E_{q2}(z,t)$.  In both cases, the incident vacuum field can be
represented by a zero, since the positive P representation yields
normally ordered expectation values as simple products.  From our
original ensemble of quadruples of wavefunctions $\boldsymbol{\psi}$
describing the atoms just before we turn on $E_s$ we obtain in this way
an ensemble of pairs $(E_{q1}(z,\tau),E_{q2}(z,\tau))$
describing the generated light field $\hat{E}_q(z,\tau)$. With this
ensemble any normally ordered field expectation value can
be found.

As an example we have investigated 
2000 atoms of mass $m$ and with 1D interaction
strengths $(g_{aa},g_{ab},g_{bb} =
(1.0,0.5,1.0)\times5\times10^{-3}\hbar\Omega l_0$ where $\Omega$ is
the frequency of the harmonic trap and $l_0=\sqrt{\hbar/m\Omega}$ is
the associated characteristic length. The atoms are spin squeezed 
by collisional interactions
for a time $t=3.0\Omega^{-1}$. They are subsequently driven towards the
state $a$, and hereafter they are illuminated with
the $E_s$ light which builds up a maximum strength of
$E_{max}=10^2\sqrt{2\pi\hbar\omega/l_0}$ in a time of roughly
$t_{rise} = 100 l_0/c$. The matter-light coupling is chosen to be
$\kappa_1=10^{-3} c/2\pi\hbar\omega$.

As we know the atoms to be spin squeezed we also know one mode of the
$\hat{E}_q$ field which will be squeezed: The one corresponding to the
simple uniform integral of $\int \psi_b^\dagger(z',0) \psi_a(z',0)
dz'$ as described in the discussion after Eq.~(\ref{eq:def_a}). Even
with the simplifications leading to Eqs.~(\ref{eq:eom_q}) and
(\ref{eq:eom_eq}) the atomic state could in principle radiate into
many other modes and do so with varying quantum statistics. To check
whether such other modes are present in this case we calculate the
first order correlation function of the field $\langle
\hat{E}_q^\dagger(L,\tau')\hat{E}_q(L,\tau)\rangle =
\overline{E_{q2}(L,\tau_2)E_{q1}(L,\tau_1)}$
($\overline{\phantom{\cdot}\ldots\phantom{\cdot}}$ indicates averaging
over the positive P ensemble). When it is diagonalized we find that
almost all population is in fact in the expected mode
$\mathcal{E}(\tau)$ which is plotted in Fig.~\ref{fig:signal}.  In
other scenarios it might be difficult to calculate beforehand exactly
which collective atomic operator is squeezed and then an analysis like
this would be necessary in order to pick the local oscillator field
for the homodyne detection.

\begin{figure}[p] 
  \begin{center} 
  \resizebox{8.5cm}{!}{
    \includegraphics{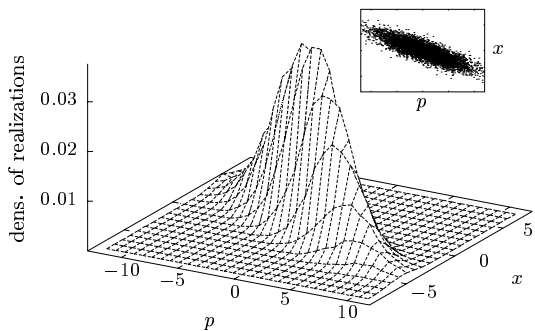}  
  }
  \caption{
  Histogram illustrating the prediction for the squeezing of
  light. The quadrature components of the output field are represented 
  by pairs of numbers Re$(a_1+a_2)/\sqrt{2}$, 
  Re$(ia_1-ia_2)/\sqrt{2}$, the distribution of which
  forms a squeezed ellipsoid shape in phase space. The actual amount
  of squeezing cannot be directly determined from this plot; it
  requires a computation of mean values and variances, making
  use of the fact that the distribution of the complex numbers $a_1$ and 
  $a_2$ represent mean values of normally ordered field operators. $10^5$
  realizations contributed to the histogram and in the insert is for
  illustration shown a scatter plot of 6800 of the representive points.
  }
  \label{fig:scatter}
  \end{center}
\end{figure}

Having confirmed that only one mode is populated we now turn to the
quantum character of the field. It depends on how we choose $\theta$
in Eq.~(\ref{eq:rot_pihalf}): For $\theta=\pi/2$ it will approximate
squeezed vacuum, for $\theta$ slightly different from $\pi/2$ it will
approximate a squeezed coherent state. As described
in~\cite{poulsen01:_posit_p_bose} the spin squeezing ellipse is at an
angle to the coordinate axes. This carries over to the light-field
which is squeezed in an appropriate quadrature component
$X_{\phi}=(\hat{a}e^{i\phi}+\hat{a}^{\dagger}e^{-i\phi})/\sqrt{2}$.  With the
above parameters, we find from the positive P simulations that the
minimum variance is $0.04\pm0.01$ corresponding to a reduction by a
factor of more than 12 from the standard quantum limit. To
illustrate the positive P results in Fig.~\ref{fig:scatter}
we show both a histogram and a scatter plot obtained from the positive
P ensemble of pairs $(a_1,a_2)$ representing
$(\hat{a},\hat{a}^\dagger)$ for the mode of the field. $a_1$ and
$a_2$, on average, are complex conjugate quantities, so that the
expectation values of hermitian field operators $\langle X_\phi\rangle
= \overline{a_1e^{i\phi}+a_2e^{-i\phi}}$ is real for all $\phi$. To
represent our simulated results we have made a histogram for the
values of $x=X_0$ and $p=X_{\pi/2}$, indicated by the {\it real part}
of $(a_1+a_2)/\sqrt{2}$ and $(ia_1-ia_2)/\sqrt{2}$. The histogram
in Fig. 2 has been obtained from $10^5$ independent realizations of
the noisy Gross-Pitaevskii equation for the atomic spin squeezing,
while the scatter plot contains only 6800 points each representing a
single realization.

The perfect output coupling of atomic correlations is the main result
of the paper, and we imagine that many other atomic quantum states may
be taken as starting point for non-classical light generation, e.g.,
EPR-correlated separated atomic ensembles, and that the mechanism may
be used also for reliable interspecies teleportation.  A squeezed
light pulse may also be applied at the "dark" input port of a beam
splitter, to cause the splitting of a strong pulse (with the
appropriate mode function ${\cal E}(t)$), into two twin pulses with
perfectly matched photon statistics.  It was recently demonstrated
that a classical light pulse can be brought to a complete halt in an
atomic sample and that it can be subsequently released \cite{lene}.
Our study suggests that a quantum light pulse can be similarly stored
and retrieved, and that one may perhaps introduce a known or unknown
signal field in a condensed sample, manipulate its quantum state by
collisional interactions and release it with the original mode
function preserved.  We emphasize that the spin squeezing mechanism is
not crucial for the squeezed light output, and for example the
QND-atomic detection and spin squeezing scheme of Kuzmich et al
\cite{kuzmich} may be an interesting possibility. Some
schemes for atomic spin squeezing may even be compatible with a
transmission or production of light with a group velocity so low, that
it may constitute a cw source of squeezed light.

\begin{thebibliography}{24}
\expandafter\ifx\csname natexlab\endcsname\relax\def\natexlab#1{#1}\fi
\expandafter\ifx\csname bibnamefont\endcsname\relax
  \def\bibnamefont#1{#1}\fi
\expandafter\ifx\csname bibfnamefont\endcsname\relax
  \def\bibfnamefont#1{#1}\fi
\expandafter\ifx\csname citenamefont\endcsname\relax
  \def\citenamefont#1{#1}\fi
\expandafter\ifx\csname url\endcsname\relax
  \def\url#1{\texttt{#1}}\fi
\expandafter\ifx\csname urlprefix\endcsname\relax\def\urlprefix{URL }\fi
\providecommand{\bibinfo}[2]{#2}
\providecommand{\eprint}[2][]{\url{#2}}

\bibitem[{\citenamefont{Fabre and Giacobino}(1992)}]{squeezing_review}
\bibinfo{editor}{\bibfnamefont{C.}~\bibnamefont{Fabre}} \bibnamefont{and}
  \bibinfo{editor}{\bibfnamefont{E.}~\bibnamefont{Giacobino}}, eds.,
  \emph{\bibinfo{title}{Quantum Noise Reduction in Optical Systems}}
  (\bibinfo{publisher}{Appl. Phys. B. {\bf 55}, 279}, \bibinfo{year}{1992}).

\bibitem[{\citenamefont{Kok et~al.}(2000)\citenamefont{Kok, Boto, Abrams,
  Williams, Braunstein, and Dowling}}]{holography}
\bibinfo{author}{\bibfnamefont{P.}~\bibnamefont{Kok}},
  \bibinfo{author}{\bibfnamefont{A.~N.} \bibnamefont{Boto}},
  \bibinfo{author}{\bibfnamefont{D.~S.} \bibnamefont{Abrams}},
  \bibinfo{author}{\bibfnamefont{C.~P.} \bibnamefont{Williams}},
  \bibinfo{author}{\bibfnamefont{S.~L.} \bibnamefont{Braunstein}},
  \bibnamefont{and} \bibinfo{author}{\bibfnamefont{J.~P.}
  \bibnamefont{Dowling}}, \emph{\bibinfo{title}{Quantum interferometric optical
  lithography:towards arbitrary two-dimensional patterns}}
  (\bibinfo{year}{2000}), \eprint{quant-ph/0011088}.

\bibitem[{\citenamefont{Bjork et~al.}(2001)\citenamefont{Bjork, Soto, and
  Soderholm}}]{bjork01:_sub}
\bibinfo{author}{\bibfnamefont{G.}~\bibnamefont{Bjork}},
  \bibinfo{author}{\bibfnamefont{L.~L.~S.} \bibnamefont{Soto}},
  \bibnamefont{and}
  \bibinfo{author}{\bibfnamefont{J.}~\bibnamefont{Soderholm}},
  \emph{\bibinfo{title}{Sub-wavelength lithography over extended areas}}
  (\bibinfo{year}{2001}), \eprint{quant-ph/0101122}.

\bibitem[{\citenamefont{Mattle et~al.}(1996)\citenamefont{Mattle, Weinfurter,
  Kwiat, and Zeilinger}}]{mattle96}
\bibinfo{author}{\bibfnamefont{K.}~\bibnamefont{Mattle}},
  \bibinfo{author}{\bibfnamefont{H.}~\bibnamefont{Weinfurter}},
  \bibinfo{author}{\bibfnamefont{P.~G.} \bibnamefont{Kwiat}}, \bibnamefont{and}
  \bibinfo{author}{\bibfnamefont{A.}~\bibnamefont{Zeilinger}},
  \bibinfo{journal}{Phys. Rev. Lett.} \textbf{\bibinfo{volume}{76}},
  \bibinfo{pages}{4656} (\bibinfo{year}{1996}).

\bibitem[{\citenamefont{Bouwmeester et~al.}(1997)\citenamefont{Bouwmeester,
  Pan, Mattle, Aibl, Weinfurter, and Zeilinger}}]{bouwmeester97}
\bibinfo{author}{\bibfnamefont{D.}~\bibnamefont{Bouwmeester}},
  \bibinfo{author}{\bibfnamefont{J.-W.} \bibnamefont{Pan}},
  \bibinfo{author}{\bibfnamefont{K.}~\bibnamefont{Mattle}},
  \bibinfo{author}{\bibfnamefont{M.}~\bibnamefont{Aibl}},
  \bibinfo{author}{\bibfnamefont{H.}~\bibnamefont{Weinfurter}},
  \bibnamefont{and}
  \bibinfo{author}{\bibfnamefont{A.}~\bibnamefont{Zeilinger}},
  \bibinfo{journal}{Nature} \textbf{\bibinfo{volume}{390}},
  \bibinfo{pages}{575} (\bibinfo{year}{1997}).

\bibitem[{\citenamefont{Furusawa et~al.}(1998)\citenamefont{Furusawa,
  S{\o}rensen, Braunstein, Fuchs, Kimble, and Polzik}}]{furusawa98}
\bibinfo{author}{\bibfnamefont{A.}~\bibnamefont{Furusawa}},
  \bibinfo{author}{\bibfnamefont{J.~L.} \bibnamefont{S{\o}rensen}},
  \bibinfo{author}{\bibfnamefont{S.~L.} \bibnamefont{Braunstein}},
  \bibinfo{author}{\bibfnamefont{C.~A.} \bibnamefont{Fuchs}},
  \bibinfo{author}{\bibfnamefont{H.~J.} \bibnamefont{Kimble}},
  \bibnamefont{and} \bibinfo{author}{\bibfnamefont{E.~S.}
  \bibnamefont{Polzik}}, \bibinfo{journal}{Science}
  \textbf{\bibinfo{volume}{282}}, \bibinfo{pages}{706} (\bibinfo{year}{1998}).

\bibitem[{\citenamefont{Hemmer et~al.}(1995)\citenamefont{Hemmer, D.~P.~Katz,
  Cronon-Golomb, Shahriar, and Kumar}}]{hemmer95}
\bibinfo{author}{\bibfnamefont{P.~R.} \bibnamefont{Hemmer}},
  \bibinfo{author}{\bibfnamefont{J.~D.} \bibnamefont{D.~P.~Katz}},
  \bibinfo{author}{\bibfnamefont{M.}~\bibnamefont{Cronon-Golomb}},
  \bibinfo{author}{\bibfnamefont{M.~S.} \bibnamefont{Shahriar}},
  \bibnamefont{and} \bibinfo{author}{\bibfnamefont{P.}~\bibnamefont{Kumar}}
  \textbf{\bibinfo{volume}{20}}, \bibinfo{pages}{982} (\bibinfo{year}{1995}).

\bibitem[{\citenamefont{Lukin et~al.}(1998)\citenamefont{Lukin, Hemmer,
  Loeffler, and Scully}}]{lukin98}
\bibinfo{author}{\bibfnamefont{M.~D.} \bibnamefont{Lukin}},
  \bibinfo{author}{\bibfnamefont{P.}~\bibnamefont{Hemmer}},
  \bibinfo{author}{\bibfnamefont{M.}~\bibnamefont{Loeffler}}, \bibnamefont{and}
  \bibinfo{author}{\bibfnamefont{M.~O.} \bibnamefont{Scully}},
  \bibinfo{journal}{Phys. Rev. Lett.} \textbf{\bibinfo{volume}{81}},
  \bibinfo{pages}{2675} (\bibinfo{year}{1998}).

\bibitem[{\citenamefont{Lukin et~al.}(1999)\citenamefont{Lukin, Matso,
  Fleischhauer, and Scully}}]{lukin99}
\bibinfo{author}{\bibfnamefont{M.~D.} \bibnamefont{Lukin}},
  \bibinfo{author}{\bibfnamefont{A.~B.} \bibnamefont{Matso}},
  \bibinfo{author}{\bibfnamefont{M.}~\bibnamefont{Fleischhauer}},
  \bibnamefont{and} \bibinfo{author}{\bibfnamefont{M.~O.}
  \bibnamefont{Scully}}, \bibinfo{journal}{Phys. Rev. Lett.}
  \textbf{\bibinfo{volume}{82}}, \bibinfo{pages}{1847} (\bibinfo{year}{1999}).

\bibitem[{\citenamefont{Wineland et~al.}(1994)\citenamefont{Wineland,
  Bollinger, and Itano}}]{wineland94:_squeez}
\bibinfo{author}{\bibfnamefont{D.~J.} \bibnamefont{Wineland}},
  \bibinfo{author}{\bibfnamefont{J.~J.} \bibnamefont{Bollinger}},
  \bibnamefont{and} \bibinfo{author}{\bibfnamefont{W.~M.} \bibnamefont{Itano}},
  \bibinfo{journal}{Phys. Rev. A}
  \textbf{\bibinfo{volume}{50}}(\bibinfo{number}{1}), \bibinfo{pages}{67}
  (\bibinfo{year}{1994}).

\bibitem[{\citenamefont{Kuzmich et~al.}(2000)\citenamefont{Kuzmich, Mandel, and
  Bigelow}}]{kuzmich}
\bibinfo{author}{\bibfnamefont{A.}~\bibnamefont{Kuzmich}},
  \bibinfo{author}{\bibfnamefont{L.}~\bibnamefont{Mandel}}, \bibnamefont{and}
  \bibinfo{author}{\bibfnamefont{N.~P.} \bibnamefont{Bigelow}},
  \bibinfo{journal}{Phys. Rev. Lett.} \textbf{\bibinfo{volume}{85}}
  (\bibinfo{year}{2000}).

\bibitem[{\citenamefont{S{\o}rensen et~al.}(2001)\citenamefont{S{\o}rensen,
  Duan, Cirac, and Zoller}}]{soerensen01:_many_bose}
\bibinfo{author}{\bibfnamefont{A.}~\bibnamefont{S{\o}rensen}},
  \bibinfo{author}{\bibfnamefont{L.-M.} \bibnamefont{Duan}},
  \bibinfo{author}{\bibfnamefont{I.}~\bibnamefont{Cirac}}, \bibnamefont{and}
  \bibinfo{author}{\bibfnamefont{P.}~\bibnamefont{Zoller}},
  \bibinfo{journal}{Nature} \textbf{\bibinfo{volume}{409}}, \bibinfo{pages}{63}
  (\bibinfo{year}{2001}).

\bibitem[{\citenamefont{Poulsen and M{\o}lmer}(2001)}]{poulsen01:_posit_p_bose}
\bibinfo{author}{\bibfnamefont{U.~V.} \bibnamefont{Poulsen}} \bibnamefont{and}
  \bibinfo{author}{\bibfnamefont{K.}~\bibnamefont{M{\o}lmer}},
  \bibinfo{journal}{Phys. Rev. A}  (\bibinfo{year}{2001}),
  \bibinfo{note}{to be published}.

\bibitem[{\citenamefont{Bouchoule and M{\o}lmer}(2001)}]{bouchoule01}
\bibinfo{author}{\bibfnamefont{I.}~\bibnamefont{Bouchoule}} \bibnamefont{and}
  \bibinfo{author}{\bibfnamefont{K.}~\bibnamefont{M{\o}lmer}}
  (\bibinfo{year}{2001}), \bibinfo{note}{to be published}.

\bibitem[{\citenamefont{Jing et~al.}(2001)\citenamefont{Jing, Chen, and
  Ge}}]{jing01}
\bibinfo{author}{\bibfnamefont{H.}~\bibnamefont{Jing}},
  \bibinfo{author}{\bibfnamefont{J.-L.} \bibnamefont{Chen}}, \bibnamefont{and}
  \bibinfo{author}{\bibfnamefont{M.-L.} \bibnamefont{Ge}},
  \bibinfo{journal}{Phys. Rev. A} \textbf{\bibinfo{volume}{63}},
  \bibinfo{pages}{15601} (\bibinfo{year}{2001}).

\bibitem[{\citenamefont{Raymer and Mostowski}(1981)}]{raymer81:_stimul_raman}
\bibinfo{author}{\bibfnamefont{M.~G.} \bibnamefont{Raymer}} \bibnamefont{and}
  \bibinfo{author}{\bibfnamefont{J.}~\bibnamefont{Mostowski}},
  \bibinfo{journal}{Phys. Rev. A}
  \textbf{\bibinfo{volume}{24}}(\bibinfo{number}{4}), \bibinfo{pages}{1980}
  (\bibinfo{year}{1981}).

\bibitem[{\citenamefont{Raymer et~al.}(1985)\citenamefont{Raymer, Walmsley,
  Mostowski, and Sobolewska}}]{raymer85:_quant_raman}
\bibinfo{author}{\bibfnamefont{M.~G.} \bibnamefont{Raymer}},
  \bibinfo{author}{\bibfnamefont{I.~A.} \bibnamefont{Walmsley}},
  \bibinfo{author}{\bibfnamefont{J.}~\bibnamefont{Mostowski}},
  \bibnamefont{and}
  \bibinfo{author}{\bibfnamefont{B.}~\bibnamefont{Sobolewska}},
  \bibinfo{journal}{Phys. Rev. A}
  \textbf{\bibinfo{volume}{32}}(\bibinfo{number}{1}), \bibinfo{pages}{332}
  (\bibinfo{year}{1985}).

\bibitem[{\citenamefont{Carman et~al.}(1970)\citenamefont{Carman, Shimizu,
  Wang, and Bloembergen}}]{carman70:_theor_stokes}
\bibinfo{author}{\bibfnamefont{R.~L.} \bibnamefont{Carman}},
  \bibinfo{author}{\bibfnamefont{F.}~\bibnamefont{Shimizu}},
  \bibinfo{author}{\bibfnamefont{C.~S.} \bibnamefont{Wang}}, \bibnamefont{and}
  \bibinfo{author}{\bibfnamefont{N.}~\bibnamefont{Bloembergen}},
  \bibinfo{journal}{Phys. Rev. A}
  \textbf{\bibinfo{volume}{2}}(\bibinfo{number}{1}), \bibinfo{pages}{60}
  (\bibinfo{year}{1970}).

\bibitem[{\citenamefont{Gardiner}(1993)}]{gardiner_cascade_prl}
\bibinfo{author}{\bibfnamefont{C.~W.} \bibnamefont{Gardiner}},
  \bibinfo{journal}{Phys. Rev. Lett.} \textbf{\bibinfo{volume}{70}},
  \bibinfo{pages}{2269} (\bibinfo{year}{1993}).

\bibitem[{\citenamefont{Gardiner and Parkins}(1994)}]{gardiner94}
\bibinfo{author}{\bibfnamefont{C.~W.} \bibnamefont{Gardiner}} \bibnamefont{and}
  \bibinfo{author}{\bibfnamefont{S.}~\bibnamefont{Parkins}},
  \bibinfo{journal}{Phys. Rev. A} \textbf{\bibinfo{volume}{50}},
  \bibinfo{pages}{1792} (\bibinfo{year}{1994}).

\bibitem[{\citenamefont{Gardiner}(1991)}]{gardiner91:_quant_noise}
\bibinfo{author}{\bibfnamefont{C.~W.} \bibnamefont{Gardiner}},
  \emph{\bibinfo{title}{Quantum Noise}} (\bibinfo{publisher}{Springer-Verlag},
  \bibinfo{address}{Berlin Heidelberg}, \bibinfo{year}{1991}).

\bibitem[{\citenamefont{Steel et~al.}(1998)\citenamefont{Steel, Olsen, Plimak,
  Drummond, Tan, Collett, Walls, and Graham}}]{steel98:_dynam_bose_einst}
\bibinfo{author}{\bibfnamefont{M.~J.} \bibnamefont{Steel}},
  \bibinfo{author}{\bibfnamefont{M.~K.} \bibnamefont{Olsen}},
  \bibinfo{author}{\bibfnamefont{L.~I.} \bibnamefont{Plimak}},
  \bibinfo{author}{\bibfnamefont{P.~D.} \bibnamefont{Drummond}},
  \bibinfo{author}{\bibfnamefont{S.~M.} \bibnamefont{Tan}},
  \bibinfo{author}{\bibfnamefont{M.~J.} \bibnamefont{Collett}},
  \bibinfo{author}{\bibfnamefont{D.~F.} \bibnamefont{Walls}}, \bibnamefont{and}
  \bibinfo{author}{\bibfnamefont{R.}~\bibnamefont{Graham}},
  \bibinfo{journal}{Phys. Rev. A}
  \textbf{\bibinfo{volume}{58}}(\bibinfo{number}{6}), \bibinfo{pages}{4824}
  (\bibinfo{year}{1998}).

\bibitem[{\citenamefont{Carusotto et~al.}(2001)\citenamefont{Carusotto, Castin,
  and Dalibard}}]{carusotto01:_n}
\bibinfo{author}{\bibfnamefont{I.}~\bibnamefont{Carusotto}},
  \bibinfo{author}{\bibfnamefont{Y.}~\bibnamefont{Castin}}, \bibnamefont{and}
  \bibinfo{author}{\bibfnamefont{J.}~\bibnamefont{Dalibard}},
  \bibinfo{journal}{Phys. Rev. A} \textbf{\bibinfo{volume}{63}},
  \bibinfo{pages}{023606} (\bibinfo{year}{2001}).

\bibitem[{\citenamefont{Liu et~al.}(2001)\citenamefont{Liu, Dutton, Behroozi,
  and Hau}}]{lene}
\bibinfo{author}{\bibfnamefont{C.}~\bibnamefont{Liu}},
  \bibinfo{author}{\bibfnamefont{Z.}~\bibnamefont{Dutton}},
  \bibinfo{author}{\bibfnamefont{C.~H.} \bibnamefont{Behroozi}},
  \bibnamefont{and} \bibinfo{author}{\bibfnamefont{L.~V.} \bibnamefont{Hau}},
  \bibinfo{journal}{Nature} \textbf{\bibinfo{volume}{409}},
  \bibinfo{pages}{490} (\bibinfo{year}{2001}).

\end{thebibliography}
\end{document}